\documentclass{raa_twocolumn}

\usepackage{float}
\usepackage{graphicx,times}             
\usepackage{natbib}
\usepackage{amssymb,amsmath}
\usepackage{caption}
\bibpunct{(}{)}{;}{a}{}{,}

\usepackage[pagebackref=true]{hyperref}

\begin{document}

  \title{GRM Scientific Pipeline
}

   \volnopage{Vol.0 (20xx) No.0, 000--000}      
   \setcounter{page}{1}          

   \author{Ping Wang 
      \inst{1*}
   \and Li Zhang \inst{1*}
   \and Jin Wang \inst{1}
   \and Wen-Hui Yu \inst{1,3}
   \and Xiao-Yun Zhao \inst{1}
   \and Shi-Jie Zheng \inst{1*}
   \and Shao-Lin Xiong \inst{1*}
   \and Yue Huang \inst{1}
   \and Jiang He \inst{1}
   \and Hao-Li Shi \inst{1}
   \and Lu Li \inst{1}
   \and Yong-Wei Dong \inst{1}   
   \and Min Gao \inst{1}
   \and Jiang-Tao Liu \inst{1}
   \and Xin Liu \inst{1}
   \and Jian-Chao Sun \inst{1}
   \and Li-Ming Song \inst{1,2}
   \and Bo-Bing Wu \inst{1}
   \and Jin-Zhou Wang \inst{1}
   \and Rui-Jie Wang \inst{1}
   \and Shuang-Nan Zhang \inst{1,2}
   }

   \institute{State Key Laboratory of Particle Astrophysics, Institute of High Energy Physics, Chinese Academy of Sciences,
             Beijing 100049, China; {\it pwang@ihep.ac.cn, zhangli@ihep.ac.cn, zhengsj@ihep.ac.cn, xiongsl@ihep.ac.cn}\\
        \and
            University of Chinese Academy of Science, Chinese Academy of Sciences, Beijing 100049, China
        \and
             School of Physics and Optoelectronics, Xiangtan University, Xiangtan 411105, China\\
\vs\no
   {\small Received 20xx month day; accepted 20xx month day}}

\abstract{ The Gamma-Ray Monitor (GRM) is a key payload of the Space-based multiband astronomical Variable Objects Monitor (SVOM) mission, which is designed to detect gamma ray bursts (GRBs) within the energy range of 15 keV to 5 MeV. The GRM Instrument Center (GRM\_IC) features real-time data processing through the X-band, enabling rapid response of high-energy GRB events. The system employs an event-driven architecture and distributed design, achieving efficient processing and real-time monitoring of massive observational data. Through comprehensive data production processes and scientific data product management, the system achieves efficient production of scientific data products of the L1B / C level through the submission of jobs to the task scheduling system. Through modular architecture design and automated processing workflow, the GRM data processing system realizes precise conversion and scientific analysis of GRB detection data, providing robust technical support for future system upgrades and cross-platform collaboration.
}

   \authorrunning{Wang, Zhang \& Wang }            
   \titlerunning{GRM Scientific Pipeline }  

   \maketitle

%
\section{Introduction}           
\label{sect:intro}

Gamma-ray bursts (GRBs) are among the most powerful explosive events in the universe, releasing immense energy across the electromagnetic spectrum in seconds to minutes. Ever since they were accidentally detected by the Vela satellite in the 1960s \citep{Klebesadel_1973}, these transient high-energy phenomena have became a central focus in astronomical research due to their cosmological origins and association with extreme physical processes, including massive star collapse and compact object mergers \citep{Woosley_1993, Eichler_1989}. The study of GRBs provides unique insights into high-energy physics, stellar evolution, and the early universe, establishing them as crucial targets for astronomical observation.

Recent advances in high-energy astronomical observation technology have significantly enhanced the sensitivity and response speed of observational instruments, enabling scientists to precisely capture transient astronomical events such as GRBs and supernovae from the distant universe. This progress has facilitated a deeper understanding of the fundamental mechanisms that govern cosmic evolution. In this context, the Space-based multiband astronomical Variable Objects Monitor (SVOM) satellite was conceived as a major mission for scientific exploration that combines international cooperation \citep{wei2016deep, 10.1117/12.2311710, 2020ChA&A..44..269Y}, forefront scientific goals, and state-of-the-art technology. SVOM aims to conduct comprehensive monitoring of transient astronomical phenomena through multi-band space observations, providing the latest scientific data and theoretical foundations for exploring high-energy explosive events in cosmic space.

The satellite integrates multiple observational instruments covering gamma-ray, X-ray, optical, and infrared wavelengths, with scientific objectives focused on GRBs, fast radio bursts (FRBs), and other transient astronomical events. In addition to conventional X-band data downlink, the satellite features real-time data transmission through the VHF band, enabling rapid response and precise positioning of high-energy burst events. This multiband, full-time-domain observation mode not only significantly enhances the efficiency of astronomical event detection, but also provides scientists with opportunities to study explosive astronomical events from multiple perspectives in real-time, thereby advancing the continuous development of high-energy astrophysics.

SVOM's scientific objectives include detecting approximately70 GRBs annually, achieving rapid and precise positioning of these events, measuring their spectral and temporal characteristics from gamma-ray to near-infrared wavelengths, and guiding timely follow-up observations by ground-based telescopes. The mission particularly focuses on increasing the detection rate of high-redshift GRBs to probe the early universe and expanding the sample of known redshift GRBs to enhance our understanding of their properties and evolution.

\begin{figure}[H]
   \centering
   \includegraphics[width=0.5\textwidth]{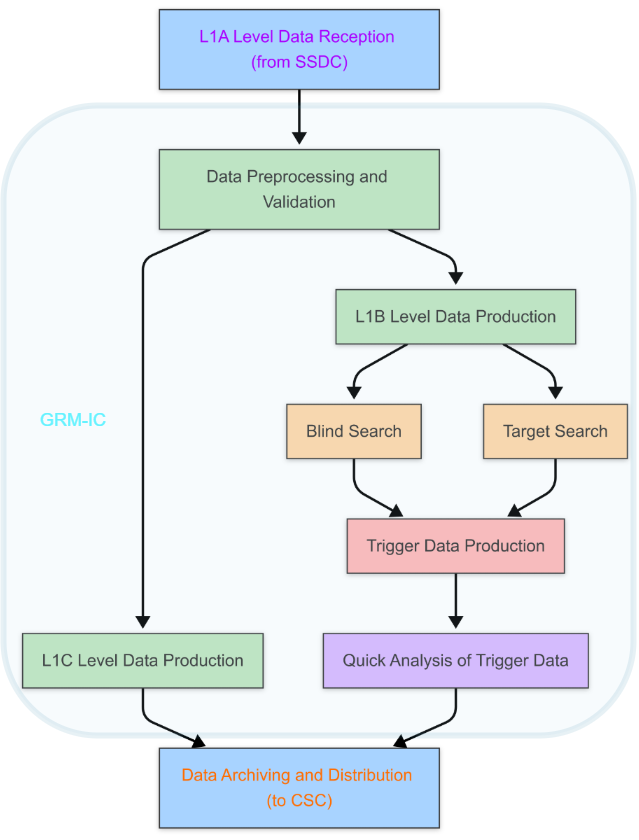}
   \caption{Logical workflow of the GRM Production Pipeline, including L1A reception, L1B/L1C production, blind/target search, trigger generation, L2 quick-look analysis, and archive/distribution.}
   \label{fig:flowchart}
\end{figure}

To achieve the production, archiving, and public distribution of standard scientific products, we developed the \textbf{GRM Scientific pipeline}. In this paper, this term denotes the complete ground system at GRM instrument center (GRM-IC), which contains two subsystems: (i) the \textbf{GRM Production Pipeline} for scientific processing (L1A $\rightarrow$ L1B (organized by hour)/L1C (organized by observation number) $\rightarrow$ L2), and (ii) the \textbf{GRM Monitoring and Scheduling System} for message reception, task scheduling, and execution monitoring. Figure~\ref{fig:flowchart} summarizes the logical flow of the GRM Production Pipeline. L1A products are preprocessed by Space Science Data Center (SSDC) and then distributed through Chinese science center (CSC); GRM-IC downloads the L1A files from CSC for downstream production. The pipeline proceeds through data validation, calibration, blind and targeted search on EVT data, automatic generation of trigger products, and quick scientific analysis of triggered events. Finally, the L1B/C data products and quick-look analysis results are archived and distributed to CSC for public access.

This paper will elaborate on each component of this data processing pipeline, including data reception and validation, categories of data products and layered production workflows, ground trigger search and quick analysis, and data archiving and distribution. The aim is to demonstrate how the GRM Instrument Center efficiently processes massive observational data and provides reliable scientific foundations for gamma-ray burst research.


\section{GRM Monitoring and Scheduling System}
\label{sect:pip}

\begin{figure*}[t]
   \centering
   \includegraphics[width=0.8\textwidth]{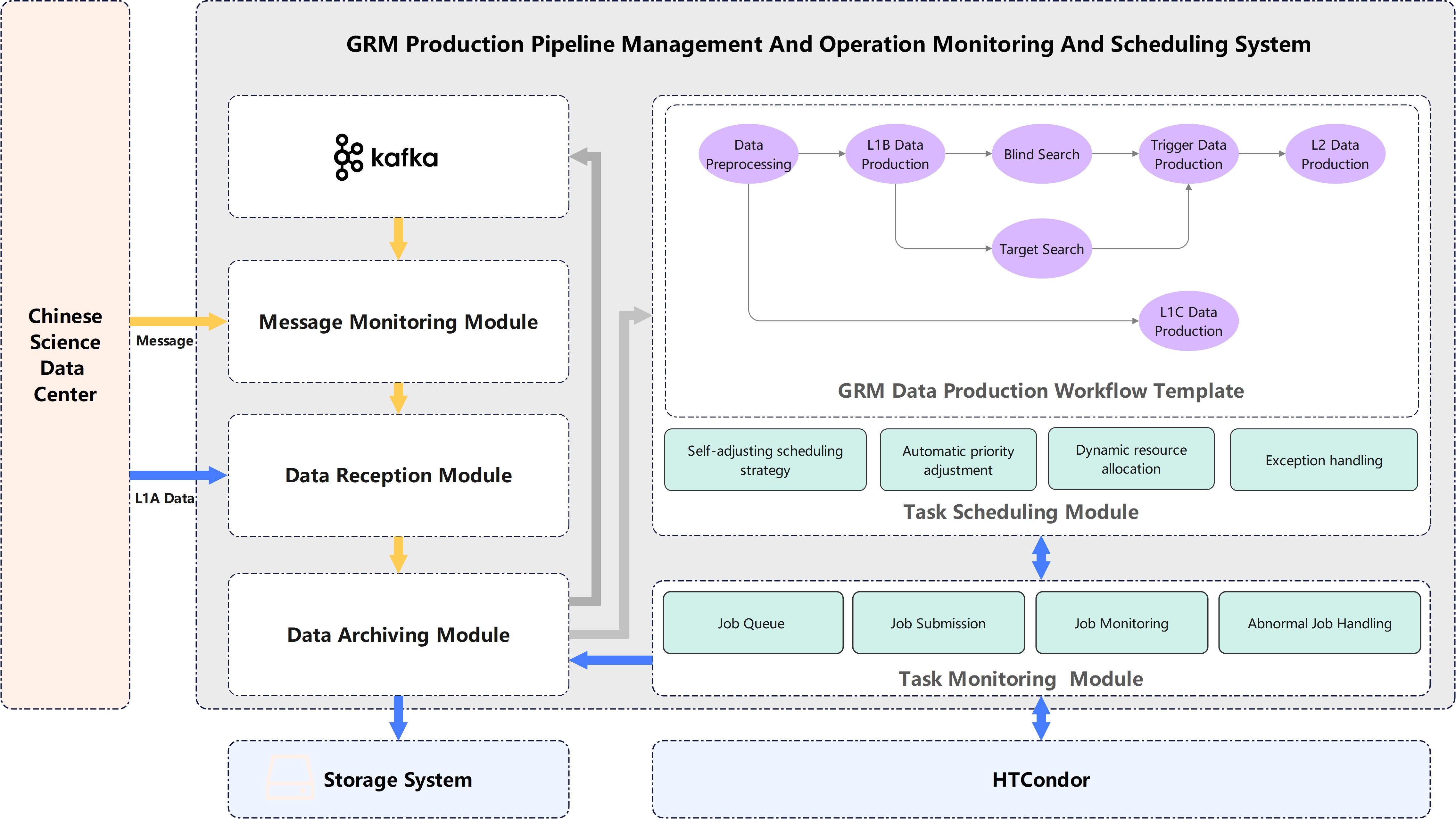}
   \caption{Architecture of the GRM Monitoring and Scheduling System, including message monitoring, data reception, data archiving, task scheduling, and task monitoring modules.}
   \label{fig:schedule}
\end{figure*}

The GRM Monitoring and Scheduling System is the control subsystem of the GRM scientific pipeline. As shown in Figure~\ref{fig:schedule}, it covers message monitoring, L1A reception, data validation/archiving, task scheduling, and job monitoring.This subsystem coordinates the production pipeline but does not replace the scientific search algorithms themselves.

(1) Message Monitoring Module:
The message monitoring module is tasked with overseeing and managing L1A-level data messages from the CSC. Basically, it handles the reception, parsing, processing, and forwarding of these messages. Using an event-driven architecture, the module ensures real-time monitoring of data messages. With multithreaded processing, it supports the concurrent reception and parsing of multiple data streams. Data transmission monitoring operates on the TCP/IP protocol and includes heartbeat detection for connection verification. Message queues facilitate decoupling from the data reception module. In addition, the module incorporates message priority classification and flow control to guarantee prioritized processing of essential scientific data.

(2) Data Reception Module:
The data reception module is responsible for receiving L1A-level data from the CSC. This module implements a memory-mapped data buffering mechanism, which supports real-time reception of high-speed data streams. Through data packet verification and integrity validation, it ensures the reliability of received data products. The module adopts a producer-consumer pattern, implementing asynchronous operations for data reception and processing, thereby improving system throughput.

(3) Data Archiving Module:
The data archiving module employs a hierarchical storage architecture, which implements classified storage and the management of data products. This module performs a quality assessment of the input data using data validation algorithms, including checks for data integrity, consistency, and validity. Through a metadata management system, it establishes comprehensive data indexing and retrieval mechanisms that support efficient data querying and access. The data archiving module also implements data version control to ensure traceability of the data product.

(4) Task Scheduling Module:
The task scheduling module plays a key role in efficiently managing data processing tasks and resources. Dynamically adjusts resource distribution and task execution order utilizing real-time monitoring data. In case of system anomalies, it optimizes execution strategies to prioritize critical data products, ensuring timely scientific data production. By employing dynamic resource allocation and priority-based queue management, it guarantees rapid processing of essential data. Task dependency analysis is utilized to optimize task sequences, enhancing system efficiency. It also supports dynamic task priority adjustments, allowing flexibility according to the system load and the importance of scientific objectives. This module is fundamental to maintaining the overall performance and stability of the data processing system.

(5) Task Monitoring Module:

The Task Monitoring Module maintains a priority queue of tasks initialized by the Task Scheduling Module. It requests resources from the HTCondor workload management system (HTCondor) according to task priority and type, and submits tasks for execution. The job monitoring component periodically scans the job logs to obtain execution status. For completed jobs, it forwards product information to the Data Archiving Module. For running jobs, it updates status information. If a task becomes abnormal or fails, the error information is returned to the Task Scheduling Module for strategy-dependent handling.

The monitoring and scheduling subsystem integrates five core modules: message monitoring, data reception, data archiving, task scheduling, and task monitoring. Through event-driven messaging and distributed execution, it improves throughput, robustness, and fault recovery for the GRM scientific pipeline.

\section{Level 1 B/C Data Production Workflow}
\label{sect:L1}

The L1B/C production module of the GRM Production Pipeline converts raw observational data into calibrated scientific products. Its modular architecture includes program initialization, task list interpretation, data processing, calibration procedures, quality assurance, and record keeping. Operating in a distributed real-time mode, it produces L1B/C products from L1A inputs through the scheduling system. The workflow begins with the L1A product as input, and then proceeds with Energy--Channel  calibration and PI transition based on the most up-to-date calibration database.

According to the data product format specifications, GRM's L1B/C level data product types are identical, with L1B level data files organized by hour and L1C level data files organized by observation number. The L1B/C level data product types include:

(1) GRD Continuous Event Data (GRM-EVT): Records time-tagged event data from three GRD detectors, with energy output in 259 PI channels (256 nominal PI channels plus 3 detector-specific super-high overflow channels, one per detector);

(2) Energy Spectrum Histogram Data (GRM-SPECHIST): aggregating counting data every 10 seconds, divided into 64 energy channels, recorded separately for high and low gain;

(3)Orbit data (GRM-ORB): Provides satellite orbital position and velocity data every second, using the J2000 and WGS-84 coordinate systems, including satellite sub-point longitude, latitude, and altitude information;

(4) Attitude data (GRM-ATT): Provides satellite attitude quaternion parameters every second;

(5) Housekeeping Data (GRM-HK): Records detector operating status every second, including temperature, high voltage state, and other operational parameters;

(6)Auxiliary Data Products (GRM-AUX): Records auxiliary parameters such as the angular relationships between the satellite, the Sun, and Earth.

\begin{figure*}[t]
   \centering
   \includegraphics[width=0.8\textwidth]{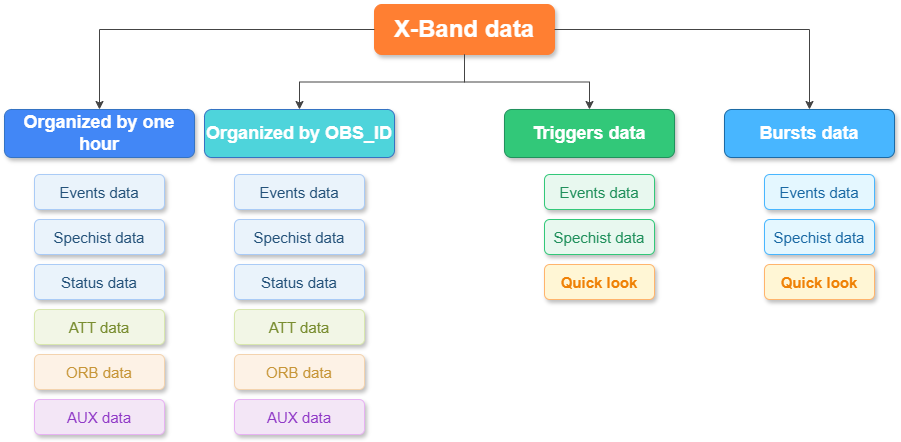}
   \caption{GRM L1B/C product family, including GRM-EVT, GRM-SPECHIST, GRM-ORB, GRM-ATT, GRM-HK, and GRM-AUX.}

   \label{fig:data_types}
\end{figure*}

The generated data product files use the FITS format standard and are archived and distributed to CSC. Figure~\ref{fig:data_types} summarizes the GRM L1B/C product family. Through modular design and automated execution, the GRM Production Pipeline provides accurate data conversion and stable scientific data delivery for SVOM.

\section{Ground Trigger Search Process}
\label{search}
As an essential component of the SVOM mission, the GRM is engineered to identify GRBs within an energy range of 15 keV to 5 MeV \citep{svom_2010}. Although real-time trigger algorithms are employed on board \citep{zhao_2013RAA}, ground-based trigger searches provide a crucial supplementary strategy for GRB detection. This approach reduces the limitations of on-board algorithms by utilizing advanced algorithms and thorough data analysis, thus significantly improving GRB detection rates, especially for weak signals and bursts with uncommon temporal or spectral properties. The ground processing pipeline incorporates a dual-mode search architecture, consisting of both blind (automatic) and targeted (manual) searches, to facilitate transient surveys of all the sky and follow-ups of external events via specialized implementations.

\subsection{Blind Search} \label{subsec:blind_search}
The blind search module for burst events is a core component of the GRM Production Pipeline for identifying transient high-energy events. The software employs a two-stage workflow to detect trigger candidates and generate standardized outputs. First, calibrated EVT data are scanned over multiple time scales, phase offsets, and energy intervals using a coherent search. Simulated templates of expected counts are compared with the observed counts in each detector channel through a log-likelihood ratio. The search is controlled by the blind-search algorithm parameter configuration file. Trigger candidates are then merged according to typical GRB durations, and qualified events are written to a trigger catalog (TCAT) file in JSON format.

\subsubsection{Data Preprocessing}
\label{subsub:data_prep}

The GRM payload consists of three detectors, each with a time resolution of 0.2 $\mu$s. Blind search scans the data at multiple temporal resolutions (from milliseconds to seconds) for each detector and across eight energy intervals. For each detector/energy-channel combination, the background is estimated from two side windows, $[t-15\,\mathrm{s},\,t-5\,\mathrm{s}]$ and $[t+5\,\mathrm{s},\,t+15\,\mathrm{s}]$, by averaging the binned counts. We note that long or complex bursts may contaminate these side windows; such candidates are re-checked in quick-look processing with adjusted background windows.

\subsubsection{Coherent Search Method}
\label{subsub:Co_search}
The coherent search method is widely performed to achieve greater sensitivity to weak signals, and is improved to search for GRBs and magnetar bursts by GECAM and GBM \cite{Blackburn_2015,Cai_2024,Cai_2021}. According to the preset shape of the gamma-ray burst energy spectrum, the particle counts of the gamma-ray bursts incident on each energy channel of each GRD detector from 3072 directions in the sky area were simulated, respectively \citep{band_1993ApJ,Cai_2024}. The Three-band spectra corresponding to spectrally hard, normal, and soft GRBs, as listed in Table~\ref{table:band_params}, are generated as the anticipated templates. The templates are then compared with the observed counts in each channel of each detector via a log-likelihood ratio.

\begin{table}[H]
    \centering
    \caption{Band spectral parameters for template generation. \label{table:band_params}}
    \begin{tabular}{lccc}
        \hline
        Spectrum Type & $\alpha$ & $\beta$ & $E_{\text{peak}}$ (keV) \\
        \hline
        Hard & 0.0 & -1.5 & 1000 \\
        Normal & -1.0 & -2.3 & 230 \\
        Soft & -1.9 & -3.7 & 70 \\
        \hline
    \end{tabular}
\end{table}

Following the presentation in \citet{Blackburn_2015} and \citet{Cai_2021}, the log-likelihood ratio LR of coherent search method is defined as:

\begin{equation}
{ \mathcal L} = \sum_{i=1}^j[{\rm ln}\frac{{\sigma}_{{n}_{i}}}{{\sigma}_{{d}_{i}}} + \frac{{\widetilde{d}_{i}}^2}{2\sigma^{2}_{{n}_{i}}} - \frac{({\widetilde{d}_{i}-{r}_{i}s)^2}}{2\sigma^{2}_{{d}_{i}}}].   \label{eq3}
\end{equation}

where $\mathcal L$ represents LR, $i$ indexes one detector--energy-channel combination, and $j$ is the total number of detector--energy-channel combinations included in the summation. ${d}_{i}$ is the observed data (counts), ${\sigma}_{{d}_{i}}$ is the standard deviation of the expected data (background+signal), ${n}_{i}$ is the estimated background, ${\sigma}_{{n}_{i}}$ is the standard deviation of the background data, $\widetilde{d}_{i}$ is the background-subtracted data, and ${r}_{i}$ and $s$ represent the instrument response and the intrinsic source amplitude, respectively.

\subsection{Target Search} \label{subsec:target_search}

The targeted search is designed for the follow-up analysis of external alerts. In the current implementation, the search time interval is defined by the input task order and then expanded by \texttt{target\_time} = 10 s on both sides, with an additional \texttt{bkg\_time} = 15 s reserved for background estimation. If alert coordinates are available, the sky search is restricted to RA $\pm 20^\circ$ and Dec $\pm 20^\circ$ around the input position. For each candidate window, a joint log-likelihood is computed by combining contributions from all active detectors and energy channels. The search uses a table of predefined algorithms with multiple time scales and phase offsets, together with several preset energy intervals.

When a possible GRB is identified via ground-based searches, the system retrieves trigger packets from the L1B data. The GRB's trigger time (T0) is set when the signal first exceeds the background threshold. Data are extracted for a period prior to T0 (usually on the order of several tens to hundreds of seconds) to characterize the background, and for the duration of the burst itself; possible afterglows lead to gathering data up to thousands of seconds post-T0. The L1B dataset is filtered to remove low-quality measurements, eliminating influences from the South Atlantic Anomaly and geomagnetic disturbances, and keeping only data from properly operating detector units. The processed data produce trigger-related outputs, including: 1. Trigger event and spechist information. 2. Light curves over multiple time scales and energy ranges. 3. A Trigger Notification containing the UTC timestamp, source position, GRB classification, initial and total flux measurements, and an estimate of event reliability.

\section{Level 2 Data Production Process}

Level 2 (L2) scientific data products are generated by burst event quick look analysis software (BEQLAS), which is automatically called by data processing pipeline\citep{zhaoyi_2018,xiong_2022,Cai_2025apj}. The pipeline, which provides task orders, initiates BEQLAS to produce L2 scientific data products. The task order must include all necessary parameters, such as input directory of scientific and auxiliary data to be processed, the temporary working directory, and the output directory for L2 data products.

The L2 data products consist of 9 kinds of products, specifically including: response matrix products for 3 GRDs DRM\_GRM\_GRDi.fits, i=1,2,3; position product PO\_GRM.fits; light curve product CLC\_GRM.fits; time lag product LAG\_GRM.fits; T90 product T90\_GRM.fits; count spectrum products for 3 GRDs CSP\_GRM\_GRDi.fits, i=1,2,3; background count spectrum products for 3 GRDs BCSP\_GRM\_GRDi.fits, i=1,2,3; spectrum product SP\_GRM.fits; hardness ratio product HR\_GRM.fits; and model count spectrum products for 3 GRDs MCSP\_GRM\_GRDi.fits, i=1,2,3. 

\begin{figure}[H]
    \centering
    \includegraphics[width=0.5\textwidth]{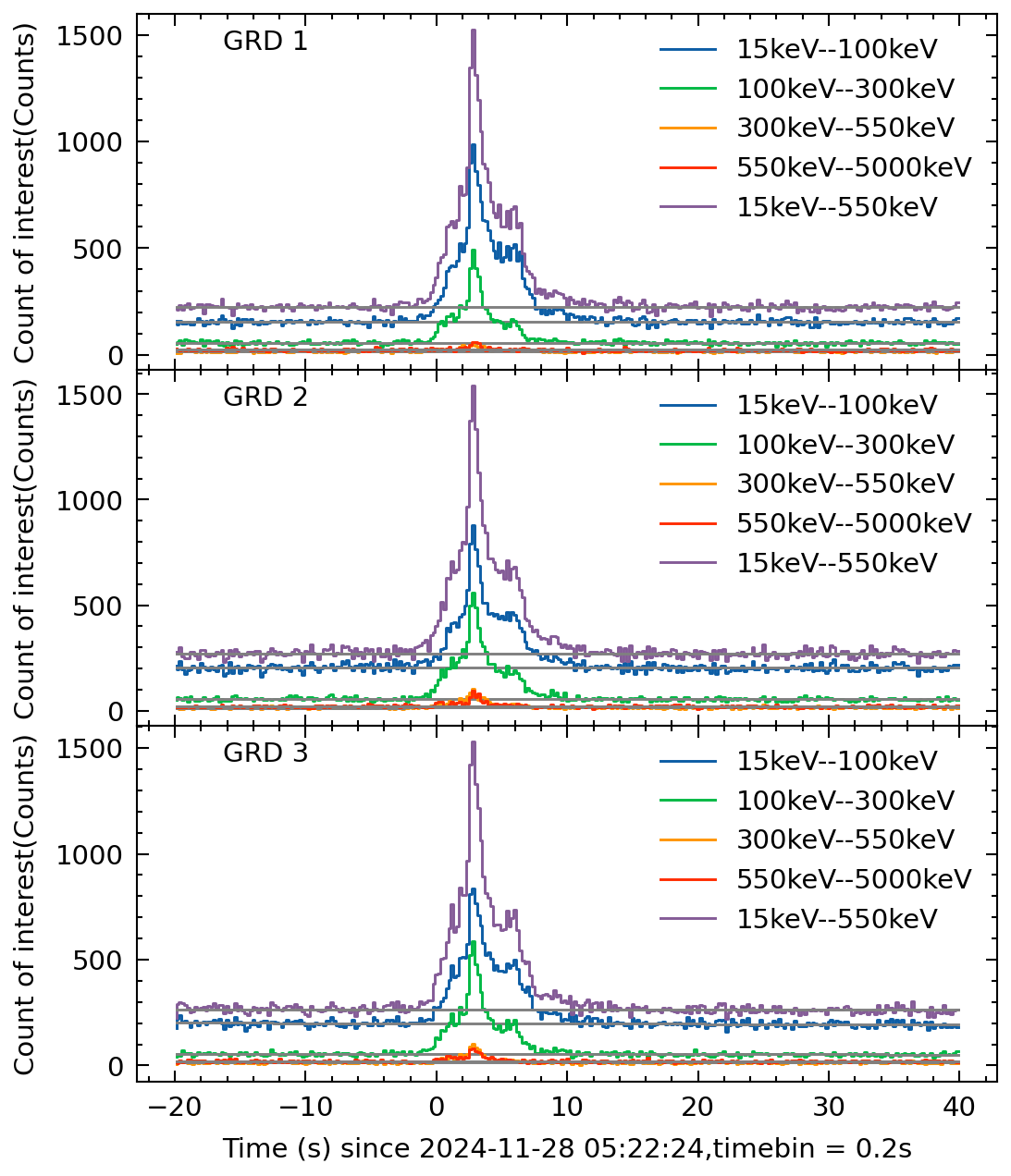}
    \caption{GRB 241128B light curve in 5 energy bands for 3 GRDs}
    \label{fig:lc}
\end{figure}

CLC\_GRM.fits stores 3 kinds of light curve corresponding to 5 different energy bands: raw count light curve (CLC), background subtracted light curve (BSCLC), background count light curve. In addition, it also stores background fitting information such as background selection interval and fitting goodness.\figurename~\ref{fig:lc} shows GRB 241128B light curve in 5 different energy bands for 3 GRDs, where gray lines are fitted background count light curve.

T90\_GRM.fits stores background subtracted cumulative count light curve (BSCCLC), which represents the accumulated photon counts from the start time. T90 and BSCCLC light curves were calculated using the CLC\_GRM.fits. To estimate the T90 error, we resample the raw light curve (Poisson) and the modeled background light curve (Gaussian) independently 1000 times, subtract the resampled background from the resampled raw light curve to obtain realizations of the background-subtracted light curve, and then derive the T90 distribution and its error from these realizations.

LAG\_GRM.fits stores time lag between light curves cooresponding to different energy bands. During propagation of a gamma-ray burst (GRB) toward Earth, photons of different energy bands may arrive at the detectors with time differences, which is defined as time lag. The time lag between these 3 energy bands (50$\sim$300 keV, 300$\sim$550 keV, 550$\sim$5000 keV) and energy band 15$\sim$50 keV are calculated. Time lag error is calculated using a method similar to that for the T90 error.

PO\_GRM.fits stores GRB position information: the right ascension (RA) and declination(DEC) coordinates of the gamma-ray burst in the J2000 celestial coordinate system, the angles between the burst source and the three main detector axes during the burst, as well as the source's position in the GRM coordinate system\citep{zhaoyi_2022}. 

DRM\_GRM\_GRDi.fits store the energy band definition of each PI channel and response matrix for each detector. Atmospheric albedo effects is considered in calculations\citep{guo_2020}. 

\begin{figure}[H]
    \centering
    \includegraphics[width=0.5\textwidth]{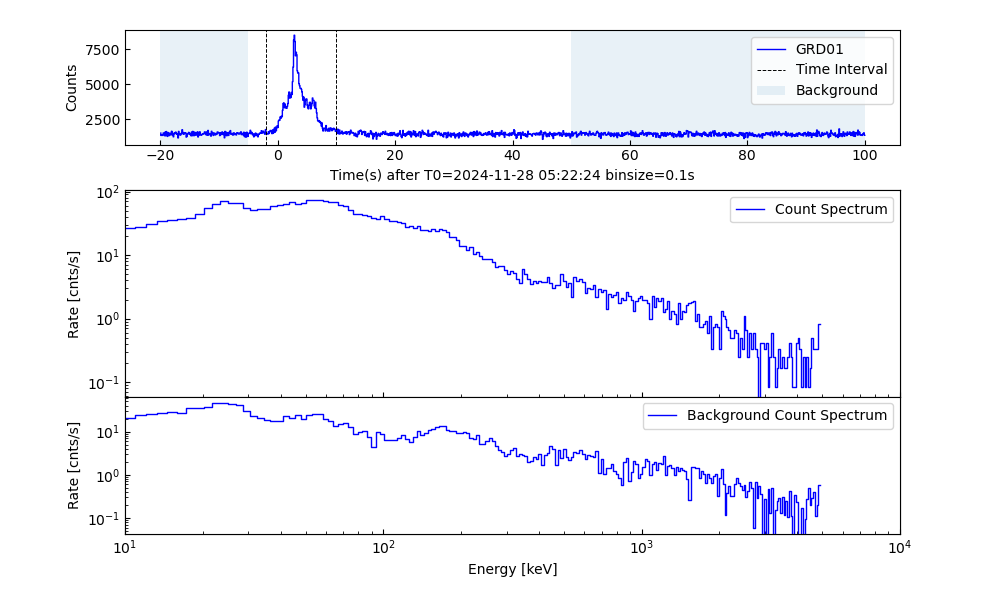}
    \caption{count spectrum and background count spectrum of GRB 241128B detected by GRD01, time interval is T$_0$+[-2.0, 10.0] s}
    \label{fig:csp}
\end{figure}

CSP\_GRM\_GRDi.fits store the Bayesian-block segmented count spectrum for each time interval. For each time slice, the background count spectrum is calculated and stored in BCSP\_GRM\_GRDi.fits. Before generating these spectra, two background intervals are required. The default intervals are $T_0+[-20.0,-2.0]$ s and $T_0+[50.0,100.0]$ s. In the current pipeline, these defaults are used as initialization and can be manually adjusted for atypical bursts (e.g., long events); a fully automated interval-optimization module is under development. The time segment between these two background intervals, $T_0+[-2.0,50.0]$ s, is divided into multiple time slices using the Bayesian-block method. For each time slice, the count spectrum and background count spectrum are generated. Figure~\ref{fig:csp} shows the count spectrum and background count spectrum of GRB 241128B detected by GRD01 during the interval $T_0+[-2.0,10.0]$ s.

SP\_GRD.fits stores spectra fitting results. For each time-resolved spectrum stored in CSP\_GRM\_GRDi.fits, three commonly used gamma-ray burst models are used to do fitting. The 3 models are powerlaw, cutoffpl, grbm. Based on the fitting results, one model is choosed as the best one. Using the best model, photon flux and fluence for 5 different energy bands are calculated and stored in SP\_GRM.fits. Peak flux is also calculated and stored.

HR\_GRM.fits stores hardness ratio for each time-resolved spectrum. The hardness ratio is the ratio of photon counts between two energy bands. The default 2 energy bands for hardness ratio calculation are: 15$\sim$100 keV for low energy band, 100$\sim$5000 keV for high energy band. For each time-resolved spectrum, three types of hardness ratios are defined: counts hardness ration (CHR), photon flux hardness ratio (PFLHR) and energy flux hardness ratio (EFLHR). CHR represents the ratio of the total counts of 3 GRDs in the high energy range to that of 3 GRDs in the low energy range. PFLHR represents the ratio of high energy photon flux to low energy photon flux. EFLHR represents the ratio of high energy photon energy flux to low energy photon energy flux. 

MCSP\_GRM\_GRDi.fits store count rate of each PI channel for each time-resolved spectrum. For each time-resolved spectrum, the count rate is derived from the convolution of the best-fit model and the detector response matrix. The effective area is taken into account during the calculation.

\section{Pipeline Performance}
\label{sect:performance}
We evaluate the performance of the \textbf{GRM scientific pipeline} for X-band data, including L1B/L1C production, blind/target search, and L2 quick-look processing. The VHF low-latency real-time alert chain is an independent system and is outside the scope of this paper.

For the GRM scientific pipeline, the mean L1A-to-L1B/C processing time for single task is below 10 minutes, and the mean trigger-search processing time is below 20 minutes. In operations, the dominant latency is upstream of pipeline execution, mainly in data reception and cross-level data conversion/transfer.

From June 26, 2024 to November 28, 2025, the GRM pipeline identified 186 high-energy transient bursts, including 177 GRBs. For GRBs, the ground blind search detected 161/177 events (91.0\%). For all transient classes combined, ground search detected 167/186 events (89.8\%).
After threshold retuning, the blind-search candidate rate decreased to about 5 per day. We report this reduction as an operational indicator of improved trigger selectivity.

\section{Summary}
\label{sect:conclusion}
We present an overview of the GRM scientific pipeline within the SVOM mission. The pipeline efficiently handles GRB data through a modular and event-driven architecture, ensuring robust processing, data quality, and system reliability.

Key features include robust data reception, structured archiving, dynamic task scheduling, and advanced ground-based trigger searches. The pipeline generates essential scientific products (L1B, L1C, and L2) for rapid transient analysis. As quantified in Section~\ref{sect:performance}, the pipeline achieves sub-10 min L1A-to-L1B/C processing and strong blind-search detection performance in current in-orbit operations. In general, the GRM pipeline effectively supports high-energy astrophysics research.

\begin{acknowledgements}

The Space-based multi-band Variable Objects Monitor
 (SVOM) is a joint Chinese-French mission led by the Chinese
 National Space Administration (CNSA), the French Space
 Agency (CNES), and the Chinese Academy of Sciences
 (CAS). We gratefully acknowledge the unwavering support of
 NSSC, IAMCAS, XIOPM, NAOC, IHEP, CNES, CEA,
 and CNRS. The authors are grateful for support from the National Key R$\&$D Program of China (grant Nos. 2024YFA1611701, 2024YFA1611700).  

\end{acknowledgements}

\bibliographystyle{raa}
\bibliography{bibtex}

\label{lastpage}

\end{document}